# TO PARALLELIZE OR NOT TO PARALLELIZE, SPEED UP ISSUE

Alaa Ismail El-Nashar

Faculty of Science, Computer Science Department, Minia University, Egypt
Assistant professor, Department of Computer Science, College of Computers and
Information Technology, Taif University, Saudi Arabia
*email:* `nashar_al@yahoo.com`

## *Abstract*

*Running parallel applications requires special and expensive processing resources to obtain the required results within a reasonable time. Before parallelizing serial applications, some analysis is recommended to be carried out to decide whether it will benefit from parallelization or not. In this paper we discuss the issue of speed up gained from parallelization using Message Passing Interface (MPI) to compromise between the overhead of parallelization cost and the gained parallel speed up. We also propose an experimental method to predict the speed up of MPI applications.*

## 1. INTRODUCTION

Execution time reduction is one of the most challenging goals of parallel programming. Theoretically, adding extra processors to a processing system leads to a smaller execution time of a program compared with its execution time using a fewer processors system or a single machine[9]. Practically, when a program is executed in parallel, the hypothesis that the parallel program will run faster is not always satisfied. If the main goal of parallelizing a serial program is to obtain a faster run then the main criterion to be considered is the speedup gained from parallelization.

Speed up is defined as the ratio of serial execution time to the parallel execution time [2], it is used to express how many times a parallel program works faster than its serial version used to solve the same problem. Many conflicting parameters such as parallel overhead, hardware architecture, programming paradigm, programming style may negatively affect the execution time of a parallel program making its execution time larger than that of the serial version and thus any parallelization gain will be lost. In order to obtain a faster parallel program, these conflicted parameters need to be well optimized.

Various parallel programming paradigms can be used to write parallel programs such as OpenMP [7], Parallel Virtual Machine (PVM) [21], and Message Passing Interface (MPI) [23]. MPI is the most commonly used paradigm in writing parallel programs since it can be employed not only within a single processing node but also across several connected ones. MPI enables the programmer to control both data distribution and process synchronization. MPICH2 [22] is an MPI implementation that is working well on a wide range of hardware platforms and also supports using of C/C++ and FORTRAN programming languages.

In this paper we discuss some of the parameters that affect the parallel programs performance as a parallelization gain issue and also propose an experimental method to predict the speed up of MPI applications. We focus on the parallel programs written by MPI paradigm using

MPICH2 implementation. This may be considered as a guide to decide before parallelizing serial applications, whether it will benefit from parallelization or not.

The paper is organized as follows: section 2 includes the related work. Section 3 covers the MPI parallelization methodologies. In section 4, we present some different parameters that affect parallel programs performance. Section 5 focuses on the performance limitations of MPI programs. In section 6, we propose an experimental method to predict the speed up of MPI programs.

## 2. RELATED WORK

Reducing program execution time is one of the advantages that application programmers hope to achieve. Converting sequential programs into parallel ones is a costly duty; it requires special hardware and software equipments. It is preferable to virtually anticipate the speed up gained from parallelism before executing the application on a real parallel environment.

Several systems have been developed for analyzing the performance of parallel programs. These systems are either model or trace based.

Petrini et al. [8] introduced a model based system to predict the performance of programs on machines prior to their construction, and to identify the causes of performance variations from the predictions. These methods pick up the slight variations in a program execution that arise at runtime that cannot be modeled by examining the static code.

Vampir [10] and Dimemas [18] are two trace based analysis tools that predict parallel programs performance. These models use a trace file and the user's selection of network parameters that is used in the communication model to simulate the program execution.

MPE (Multi-Processing Environment) library and jumpshot [1] that are distributed with MPICH [22] implementation provide graphical performance analysis for message passing interface programs.

In this paper we introduce an experimental approach to predict the speed up of message passing programs. Our approach is based on executing the parallel program several times on a single physical processor with different numbers of virtual MPI processes.

## 3. PARALLELIZATION WITH MPI

In message passing paradigm, several separate processes used to complete the overall computation. In this scheme, many concurrent processes are created, and all of the data involved in the calculation is distributed among them using different ways. There is no shared data; when a process needs data held by another one, the second process must send it to the first process. An MPI message passing protocol describes the internal methods and policies an MPI implementation employs to accomplish message delivery. There are two common message passing protocols, eager and rendezvous [13], [17]. Eager protocol is an asynchronous protocol that allows a send operation to complete without acknowledgement from a matching receive. Rendezvous protocol is a synchronous protocol which requires an acknowledgement from a matching receive in order to complete the send operation. Since MPI enables the programmer to control both data distribution and process synchronization, problem decomposition and inter process communication represent two challenges in writing MPI parallel programs. Unless they are coded carefully, program performance will be negatively affected.

### 3.1 Problem decomposition

The first challenge in writing MPI programs is how to divide the concerned problem into smaller sub problems. Problem decomposition has two types, data parallelism and task parallelism.

Data partitioning challenge concerns with the manner in which the data can be divided among the available processors. Data are divided into pieces of approximately the same size and then mapped to different processors or MPI processes depending on the process ID. Each processor/process then operates only on the portion of the data that is assigned to it. This strategy can be efficiently used in solving the iterative problems in which processors can operate independently on large portions of data, communicating only the much smaller data pieces at each iteration. The processes may need to communicate periodically in order to exchange data. This approach implies that the program needs to keep track of date pieces required by a process at any time instance.

Task parallelism focuses on the computation that is to be performed rather than on the data manipulated by the computation. The problem is decomposed according to the work that must be done. Each task then performs a portion of the overall work.

### 3.2 Processes Communication

Inter process communication challenge concerns with the manner in which the running processes can be fully controlled. This implies that explicit send and receive data operations must be executed whenever data needs to move from one process to another. Two approaches can be used to implement data distribution and communication activities among processors, namely, "point-to-point communication" and "collective communication".

### 3.2.1 Point-to-Point Communication

MPI point-to-point operations enable message passing between only two different MPI processes. In this scheme, one process performs the message send operation and the other one performs the matching receive operation. Send and receive operations work in two modes, blocking and non-blocking. In blocking mode, A blocking send routine will only return after it is safe to modify the application send buffer. This implies a handshaking with the receive process to confirm a safe send. A blocking receive only returns after the data has arrived and is ready for use by the program. In case of  non-blocking mode, both send and receive routines return immediately and do not wait for any communication events to complete, such as message copying from user memory to system buffer space or the actual arrival of message. In this mode, non-blocking operations request the MPI library to perform the operation when it is able. It is unsafe to modify the application buffer until the requested non-blocking operation was actually performed by the library. There are "wait" routines used to do this task. Non-blocking communications are primarily used to overlap computation with communication and exploit possible performance gains [20].

### 3.2.2 Collective Communication

In general, all data movement among processes can be accomplished using MPI send and receive routines. More over, a set of standard collective communication routines [20] are defined in MPI. Each collective communication routine has a parameter called a communicator, which identifies the group of participating processes. The collective communication routines allow data movement among all processors or just a specified set of processors.

The function MPI_Barrier blocks the caller processes until all members in the communicator call the routine. The function MPI_Bcast broadcasts a message from the root process to all processes in the communicator. The routines MPI_Gather and MPI_Gatherv allow each process in the communicator to send data to one process, while MPI_Scatter allows one process to send a different message to each other process. The routines MPI_ Allgather and MPI_Allgatherv gather fixed and variable sized information, respectively, from all processes and puts the results to all processes. The function MPI_Alltoall is a generalization of MPI_Allgather, it allows different messages to be sent to different processes. The most general form of all-to-all communication is MPI_Alltoallv, which allows general many-to-many or one-to-many communications to be performed by carefully selecting the input arguments. Finally, MPI_Reduce performs global reduction operations using an operation, such as sum, maximum or minimum, which is then sent to the root process [4].

## 4. PERFORMANCE METRICS

Three metrics are commonly used to measure the performance of MPI programs, execution time, speedup and efficiency. Several factors such as the number of processors used, the size of the data being processed and inter-processor communications influence parallel program's performance

### 4.1 Execution time

A parallel program's execution time is a common performance indicator. It is defined as the time elapsed from that instance at which the first processor starts program execution to that instance at which the last processor completes it. MPI enables the programmer to measure the execution time of his code or a part of it by calling the function MPI_wtime( ). This function call returns the wall clock time in seconds represented as double precision value on the calling processor. The part of code to be timed is enclosed between two timing calls, the difference between the two time values that generated from timing calls is the execution time of this part of code. The execution time $T$ is given by:

$$T = T_{Comp} + T_{Comm} + T_{idle} \qquad (1)$$

where $T_{Comp}$ is the computation time, $T_{Comm}$ is the communication time consumed by processor to send and/or receive messages, and $T_{idle}$ is the time a process spends waiting for data from other processors.

### 4.2 Speed up and Efficiency

Considering execution time only as a performance metric may be insufficient, specially if we need to study how the number of processors and problem size can affect a program performance.

Speed up is another performance metric that takes processors number p, and problem size n, into account. In terms of problem size and processors number, the total parallel execution time of a program that solves an n size problem on p processors is given by:

$$T_{parallel} = \sigma(n) + \frac{\varphi(n)}{p} + \kappa(n, p) \qquad (2)$$

Where $\sigma(n)$ is the program's serial part execution time, $\varphi(n)$ is the program's parallel part execution time, and $\kappa(n, p)$ is the communication time.

Speed up is the ratio of the time taken to solve a problem on a single processor to the time required to solve the same problem on a parallel computer with multiple processors [24]. The speedup metric for solving an n-size problem using P processors is expressed by:

$$\psi(n, p) \le \frac{T_{serial}}{T_{parallel}} \qquad (3)$$

Amdahl's Law [7] is one way of predicting the maximum achievable speedup for a given program. The law assumes that a fraction f of a program's execution time was infinitely parallelizable with no overhead, while the remaining fraction, 1-f, was totally serial [15]. According to this law, the speedup of n-size problem on p processors is governed by

$$\psi(n, p) \le \frac{1}{f + (1-f)/p} \ , \ 0 \le f \le 1 \qquad (4)$$

Amdahl's law treats problem size as a constant and hence the execution time decreases as number of processors increases. Gustafson law [12] gives another formula for predicting maximum achievable speedup which is described by

$$\psi(n, p) \leq p + (1 - p)s \qquad (5)$$

where s is the fraction of total execution time spent in serial code. The two laws ignore the communication cost ; they overestimate the speed up value [3].

Efficiency is the ratio of speed up obtained to the number of processors used [2]. It measures processors utilization. Parallel system efficiency of solving an n-size problem on P processors is given by

$$0 \leq \varepsilon(n, p) \leq \frac{\psi(n, p)}{p} \leq 1 \qquad (6)$$

# 5. PERFORMANCE LIMITATIONS OF MPI PROGRAMS

Several factors affect the performance of parallel MPI programs. The application programmer have to adapt these variables to achieve the optimal performance.

## 5.1 Effect of problem decomposition

When dividing the data into processes the programmer have to pay attention to the amount of load being processed by each processor. Load balancing is the task of equally dividing work among the available processes. This is easy to be programmed when the same operations are being performed by all the processes on different pieces of data. Irregular load distribution leads to load imbalance which cause some processes to finish earlier than others. Load imbalance is one source of overhead, so all tasks should be mapped onto processes as evenly as possible so that all tasks complete in the shortest amount of time to minimize the processors' idle time which lead to a faster execution as equation 1 indicates.

## 5.2 Effect of communication pattern

The cost of communication in the execution time can be measured in terms of latency and bandwidth. Latency is the time taken to set up the envelope for communication, where bandwidth is the actual speed of transmission. Regardless of the network hardware architecture the communication pattern affects the performance of MPI programs. Using collective communication pattern is more efficient than using of point-to-point communication pattern [23], so the application programmer have to avoid using of the latter one as much as possible, specially for large size problems, for the following reasons:

1. Although point-to-point pattern is a simple way of specifying communication in parallel programs; its use leads to large program size and complicated communication structure, which negatively affect the program performance.
2. Send-receive does not offer fundamental performance advantages over collective operations. The latter offer efficient implementations without changing the applications.
3. In practice, using the non-blocking versions of send-receive, MPI_Isend and MPI_Irecv, often lead to slower execution than the blocking version because of the extra synchronization.

## 5.3 Effect of message size

Message size can be a very significant contributor to MPI application performance. The effect of message size is also influenced by latency, communication pattern and number of processors used as described in equation 2 and equation 3. To achieve an optimal performance, the application programmer should take the following considerations into account:

1. In most cases, increasing the message size will yield better performance. For communication intensive applications, the smaller message size reduces MPI application performance because latency badly affects short messages..
2. for smaller message size with less number of processors, it is better to implement broadcasting in terms of non-blocking point-to-point communication whereas for other cases broadcasting using MPI_Bcast saves time significantly.

## 5.4 Effect of message passing protocol

MPI message passing protocols affect the program performance. The performance is implementation dependent. So the application programmer has to consider the following circumstances:

1. In case of eager protocol, the receiving process is responsible for buffering the message upon its arrival, specially if the receive operation has not been posted [13]. This operation is based upon the implementation's guarantee of a certain amount of available buffer space on the receive process. In this case, the application programmer has to pay attention to the following requirements to achieve a reasonable performance
   a. message sizes must be small.
   b. avoid using of intensive communication to decrease the time consumed by the receive process side to pull messages from the network and/or copy the data into buffer space.
2. If the receiving process buffer space can't be allocated or the limits of the buffer are exceeded rendezvous protocol is used. In this protocol, sender process sends message envelope to destination process which receives and stores that envelope. When buffer space is available, destination process replies to sender that requested data can be sent, hence sender process receives reply from destination process and then sends data [17]. In this case, the application programmer has to pay attention to the following requirements to achieve a reasonable performance
   a. message sizes must be large enough to avoid the time consumed for handshaking between sender and receiver.
   b. Using non-blocking sends with waits/tests to prevent program from blocking while waiting for a receiving confirmation from receive process.

## 5.5 Effect of processors' number

Adding extra processors to the system reduces the computation time but increases the communication time as described in equation 3. The increase in communication time may be larger than the decrease in computation time which leads to a dramatic decreasing of performance. Equation 4 assures that the speedup is usually less than the number of processors. In practice, speed up does not increase linearly as the number of processors increases but tends to saturate and accordingly the efficiency drops as the number of processors increases [12].

The effect of processor's number is also influenced by the problem size. Speedup and efficiency increase as the problem size increases on the same number of processors. If increasing the number of processors reduces efficiency, and increasing the problem size increases efficiency, the application programmer should be able to keep efficiency constant by increasing both simultaneously.

## 5.6 Effect of processes' number

MPI implementations allow the programmer to run his application using arbitrary number of processes and processors. The number of processes may be less than, equal to, or greater than the number of processors. It is common to develop parallel applications with a small number of processes on a single processor. As the application becomes more fully developed and stable, larger testing runs can be conducted on actual clusters to check for scalability and performance bottlenecks.

The number of processes per processor affects the application performance so the application programmer has to be aware of the following considerations:

1. In general, maximum performance is achieved when each process has its own processor. When the number of processes is less than or equal to the number of processors, the application will run at its peak performance. Since the total system is either underutilized (there are unused processors) or fully utilized (all processors are being used), the application is not hindered by several parameters such as context switching, cache misses, or virtual memory thrashing caused by other local processes [14].

2. running too many processes, the processors will thrash, continually trying to give each process its fair share of run time.

3. running too few processes may not enable the programmer to run meaningful data through his application, or may not cause error conditions that occur with larger numbers of processes.

## 6. EXPERIMENTAL SPEED UP PREDICTION

In some cases, the predicted performance may differs from that achieved experimentally. In this section we present an experimental method to predict the speed up of MPI applications as a performance measure. The proposed method is summarized in the following steps:

1. Execute the serial version of  MPI application on a single processor machine.

2. Record the serial execution time, $T_s$.

3. Execute the parallel MPI application on the same single processor machine repeatedly using arbitrary number of MPI processes, 1,2,3,…,n.

4. Record the parallel execution times, $Tp_1, Tp_2, ...., Tp_n$, for each run.

5. Graph the obtained results as a two dimensional graph. The X-axis for MPI processes number and the Y-axis for the parallel execution times, $Tp_1, Tp_2, ...., Tp_n$.

6. If the parallel execution time is rapidly increases as the number of MPI processes increases, this implies that the MPI application will exhibit a poor speed up if it is run in parallel on multiple physical processors.

7. If the parallel execution time remains constant or  slowly increases as the number of MPI processes increases, this implies that the MPI application will exhibit a linear speed up if it is run in parallel on multiple physical processors.

We applied the proposed method on two MPI applications. The first one solves the concurrent wave equation and  the second finds the number of primes and also the largest prime number within an interval of integers. The two applications are also executed in parallel on multiple physical processors. The recorded serial execution time, $T_s$ for both applications is used to find out their experimental speed up to be compared with the predicted ones.

### 6.1 Experimental setup

Since modern parallel machines are very costly and not easy to be access, we used an experimental system consists of 8 DELL machines. Each of these machines consists of  Intel i386 based  P4-1.6GHz processor with 512MB memory running on Microsoft Windows XP Professional Service Pack 2. These machines are connected via a Fast Ethernet 100Mbps switch. These machines are not as powerful as the recent cluster machines in terms of the hardware and  performance but they can reasonably perform for testing purposes and also for solving small and middle size parallel problems. The experiments programs was written in Fortran 90 using MPICH2 version 1.0.6p1, as a message passing implementation.

## 6.2 Experimented Problems

### 6.2.1 Problem 1: Concurrent wave equation

The concurrent wave equation [6] is the partial differential equation that describes the propagation of waves. The one-dimensional wave equation that represents a flexible vibrating string stretched between two points on the x-axis is expressed by $\frac{\partial^2 u}{\partial t^2} = c^2 \frac{\partial^2 u}{\partial x^2}$, where, c is the speed of the wave's propagation and $u = u(\vec{p}, t)$ describes the wave's amplitude at position $\vec{p}$ at time $t$.

The numerical solution of this equation can be given by :

$$u(i,t+1) = (2.0 * u(i,t)) - u(i,t-1) + (c *(u(i-1,t)-(2.0*u(i,t))+u(i+1,t))) \qquad (8)$$

where i is the position index along the x axis at the time t. Equation 8 implies that the amplitude at each position index i and time t+1 depends on the previous time steps (t, t-1) and neighboring points (i-1, i+1).This means that the parallel solution requires interprocess communication. The parallel solution is based on  dividing the vibrating string into points. Each processor is repeatedly responsible for updating the amplitude of a number of points over time. At each iteration, each processor exchanges boundary points with their nearest neighbors. The parallel algorithm that solve this equation is summarized as follows:

```
 1. Initialize MPI environment.
 2. Determine number of MPI processes and identities.
 3. Determine left and right neighbors.
 4. If  Process_id=master then
 5.      obtains input values from user.
 6.      broadcast time advance parameter, total points and time steps
 7. else
 8.      receive input values from master
 9. endif
10.calculate initial values based on sine curve
11.  calculate new values using wave equation
12.  update their points a specified number of times
13.  update values for each point along string
14.  exchange data with "left-hand" neighbor
15.  exchange data with "right-hand" neighbor
16.  If  Process_id <> master then
17.      send the updated values to the master
18.  else
19.      receives results from workers and prints
20.  endif
21. Finalize MPI environment
22. End
```

Figure 1. Wave equation parallel algorithm solution

### 6.2.2 Problem 2: Prime numbers generator

There is no general "formula" for generating prime numbers. However there are some approximations and theorems predicting the number of prime numbers less than a particular upper bound [11]. Brute-force algorithm [5], shown in figure 2, which is also called "naïve" algorithm can be used in primality test .

```
 1. Naïve (n:integer,prime:logical)
 2. /*  Assume first four primes are counted
 3. if n > 10 then
 4.     squareroot = int( √n )
 5.     do i=3,squareroot,2
 6.         if n % i = 0 then
 7.             prime = false
 8.             return
 9.         endif
10.     enddo
11.     prime = true
12.     return
13. else
14.     prime =false
15.     return
16. endif
17. End naïve
```

Figure 2. Naïve and Sieves algorithm

The simplest primality test for a given number n, is to check whether any integer m from 2 to $n - 1$ divides n. If n is divisible by any m then n is composite, otherwise it is prime. Rather than testing all m up to $n - 1$ , "naïve and sieves" algorithm [16] tests only  m up to $\sqrt{n}$ , if n is composite then it can be factored into two values, at least one of which must be less than or equal to $\sqrt{n}$ . The algorithm efficiency can also be improved by skipping all even m except 2.

A pseudo serial version and also the corresponding MPI parallel version that use this algorithm to find the number of primes and also the largest prime number within an interval of integers are shown in figure 3 and figure 4 respectively.

```
 1. Determine the upper LIMIT of integers interval.
 2. prime_counter = 4
 3. do n =11, LIMIT, 2
 4.     call naïve(n, prime)
 5.       if  (prime) then
 6.             prime_counter = prime_counter + 1
 7.             prime_value = n
 8.       endif
 9.   Enddo
10. print  prime_value, prime_counter
11. End
```

Figure 3. Serial primes generator pseudo code.

```
1. Initialize MPI environment.
2. Determine number of MPI processes, ntasks, and Identities, rank.
3. Determine the upper LIMIT of integers interval.
4.  mystart = (rank*2) + 1; stride = ntasks*2
5. prime_counter = 0; prime_value = 0
6. do n=mystart, LIMIT, stride
7.        call naïve(n, prime)
8.        if  (prime) then
9.             prime_counter = prime_counter + 1
10.            prime_value = n
11.       endif
12. enddo
13. Reduce(pc,pcsum,MPI_SUM,master)
14. Reduce(prime_value, maxprime, MPI_MAX , master)
15. If  Process_id=master then
16.     print  maxprime,pcsum-4
17.  endif
18. Finalize MPI environment
19. End
```

Figure 4. Parallel MPI primes generator pseudo code.

## 6.3 Predicted versus experimental results

The parallel MPI applications that solve both wave equation and prime numbers generator problems were executed on the hardware architecture described in section 5.1. Serial execution time , parallel execution time on a single processor using multiple number of processes and also parallel execution time on multiple processors for both problems are shown in table 1.

Table 1. Serial and parallel execution times for
Wave Equation and Primes Generator

| Problem | Serial execution time | Parallel execution | | | |
|---|---|---|---|---|---|
| | | Single physical processor | | Multiple physical processors | |
| | | MPI processes | Execution time | Physical processors | Execution time |
| Problem 1 Wave Equation | 0.80216 | 1 | 1.3561 | 1 | 1.3561 |
| | | 2 | 3.6942 | 2 | 4.0952 |
| | | 3 | 6.3833 | 4 | 1.2112 |
| | | 4 | 9.4002 | 8 | 11.4501 |
| | | 5 | 12.5629 | | |
| | | 6 | 15.301 | | |
| | | 7 | 18.1778 | | |
| | | 8 | 21.5001 | | |
| | | 9 | 24.1733 | | |
| | | 10 | 27.3349 | | |
| Problem 2 Primes Generator | 55.625 | 2 | 55.5887 | 1 | 57.625 |
| | | 4 | 55.464 | 2 | 32.6704 |
| | | 8 | 54.9653 | 4 | 17.38331 |
| | | 10 | 55.5158 | 6 | 11.58861 |
| | | 16 | 55.1428 | 8 | 8.2103 |
| | | 20 | 55.9213 | | |

Applying the proposed speed up prediction method to wave equation problem using 10 MPI processes on a single physical processor we predicted that the application will exhibit a poor speed up if it is executed in parallel using multiple physical processors.

Our prediction is based on that the execution time is rapidly increases as the number of MPI processes as shown in figure 5. To prove that our prediction was true, we executed the same MPI code on 8 physical processors. Knowing the execution time of the serial code version, the experimental speed up was calculated. Figure 6 shows that the maximum speed up achieved by 8 physical processors was only 0.66228534 and hence our prediction was true.

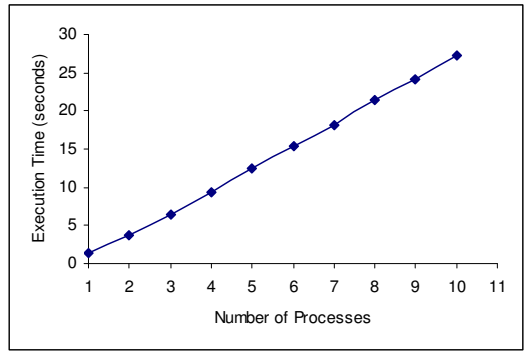

Figure 5. Execution time using 10 processes on a single CPU for problem 1

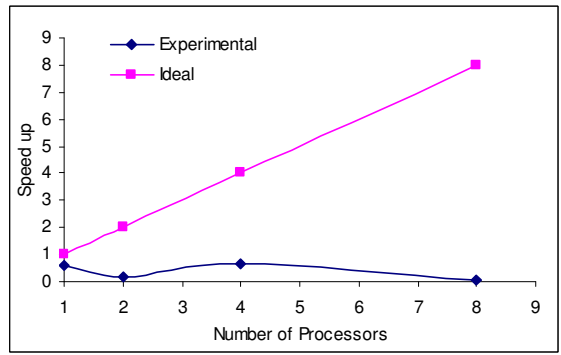

Figure 6. Experimental speed up for problem 1

To be unbiased, we also re-executed the same parallel code using different number of processes on the same 8 physical processors. Figure 7 shows that the execution time was negatively affected as the number of MPI processes increases except in case of running a small number of MPI processes using 8 physical processors. The experimental results shows that there is no significant speed up improvement as shown in figure 8. This also proves that our prediction was true.

Applying the proposed method to prime numbers generator problem using 20 MPI processes on a single physical processor, we predicted that the application will exhibit a linear speed up if it is executed in parallel using multiple physical processors.

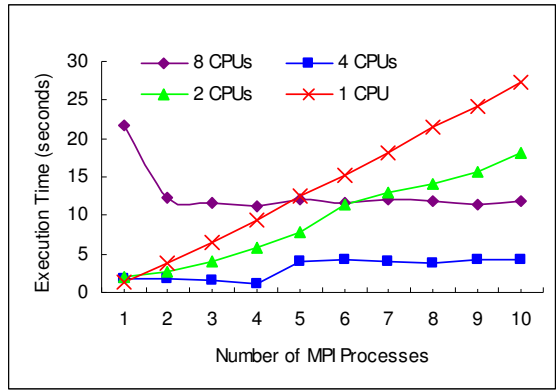

Figure 7.   Effect of  processes number on execution time using 8 CPUs  for problem1

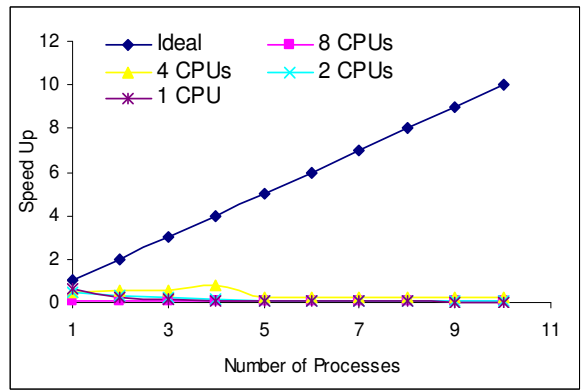

Figure 8. Experimental vs. ideal speed up for problem 1

Our prediction is based on that the execution time is slowly increases or seems to be constant as the number of MPI processes as shown in figure 9. Running the same MPI code on 8 physical processors achieved a linear speed up as shown figure 10 and hence our prediction was also true.

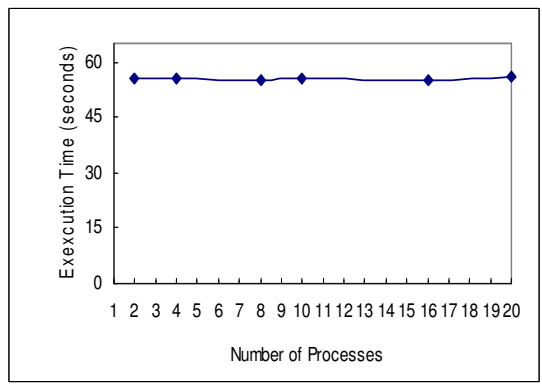

Figure 9. Execution time using 20 processes on a single CPU  for problem 2

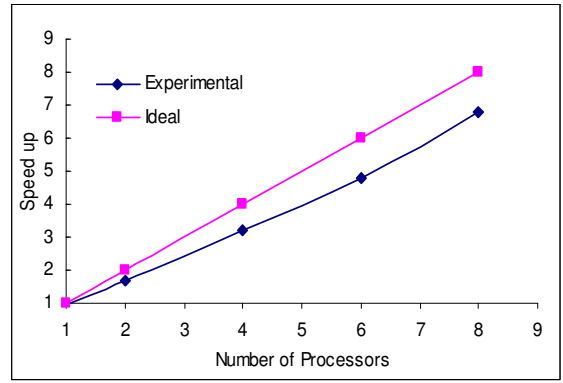

Figure 10. Experimental speed up for problem2

## 7. CONCLUSION

Concerning the issue of speed up gained from parallelization, the decision making to parallelize or not to parallelize the serial application is not a trivial task.

In this paper we studied the conflicting parameters that affect the parallel programs performance, specially MPI applications, showing some recommendations to be followed to achieve a reasonable performance. The problem nature is one of the most important factors that affect the parallel program speed up. If The problem can be divided into independent subparts and no communication is required, except to split up the problem and combine the final results, then there is a great parallelization opportunity, and the resultant parallel program will exhibit a linear speed up. If the same instruction set are applied to all data and processes communication is synchronous , speed up will be directly proportional to the computation -communication ratio. If there are different instruction sets to be applied to all data to solve a specific problem and the inter-process communication is asynchronous, this will reduce the parallelization opportunity. Speed up of the resultant parallel application will be negatively affected with extra communication overhead.

We also proposed an experimental method that aids in speed up prediction. The proposed method is based on running the MPI applications with several MPI processes using only one single processor machine. It gives an indication about the speed up behavior of MPI applications without using extra parallel hardware facilities, so it is recommended to be applied to MPI applications before running them on real powerful cluster machines or an expensive parallel systems. The proposed method was applied to predict the speed up of MPI applications that solve wave equation and prime numbers generator problems. The predicted speed up was as the same as experimental speed up achieved when using multiple physical processors for both applications.

## REFERENCES

[1]  A. Chan D. Ashton, R. Lusk, and W. Gropp, Jumpshot-4 Users Guide,  Mathematics and Computer Science Division, Argonne National Laboratory July 11, 2007.

[2] A. Grama, A. Gupta, and V. Kumar, "Isoefficiency Function: A Scalability Metric for Parallel Algorithms and Architectures", IEEE Parallel and Distributed Technology, Special Issue on Parallel and Distributed Systems: From Theory to Practice, Volume 1, Number 3, pp 12-21, August 1993.

[3] A. H. Karp and H. Flatt, "Measuring Parallel Processor Performance", Communication of the ACM Volume 33 Number 5, May 1990.

[4] A. Karwande, X. Yuan, and D. K. Lowenthal, " CC-MPI: A Compiled Communication Capable MPI Prototype for Ethernet Switched Clusters", Journal of Parallel and Distributed Computing, Volume 65, Number 10, pp 1123-1133, 2005.

[5] A. Mohammad , O. Saleh and R. A. Abdeen "Occurrences Algorithm for String Searching Based on Brute-force Algorithm", Journal of Computer Science, Volume 2, Number 1, pp 82-85, 2006.

[6] C. Geoffrey, Fox et al "Solving problems on concurrent processors", Prentice-Hall, Inc. Upper Saddle River, NJ, USA, ISBN:0-13-823022-6 , 1988.

[7] E. Gabriel, G. E. Fagg, G. Bosilca, T. Angskun, J. J. Dongarra, J. M. Squyres, V. Sahay, P. Kambadur, B. Barrett, A. Lumsdaine, R. H. Castain, D. J. Daniel, R. L. Graham, and T. S. Woodall, "Open MPI: Goals, Concept, and Design of a Next Generation MPI Implementation", In Proceedings, 11th European PVM/MPI Users' Group Meeting, Budapest, Hungary, pp. 97–104, September 2004.

[8] F. Petrini, D. Kerbyson, and S. Pakin. The case of the missing supercomputer performance: Achieving optimal performance on the 8,192 processors of ASCI Q. In Proc. Supercomputing, Phoenix, AZ, Nov. 2003.

[9] G. M. Amdahl, "Validity of the Single Processor Approach to achieving Large Scale Computing Capabilities", In Proceedings of the AFIPS Spring Joint Computer Conference, pp 483–485, April 1967.

[10] H. Brunst, M. Winkler, W. E. Nagel and H.-C. Hoppe, Performance Optimization for Large Scale Computing: The Scalable VAMPIR Approach,, International Conference on Computational Science (ICCS2001) Workshop on Tools and Environments for Parallel and Distributed Programming, San Francisco, CA, May 2001.

[11] I. Aziz, N. Haron, L. Tanjung and W. W. dagang, "Parallelization of Prime Number Generation Using Message Passing Interface", WSEAS Transactions on Computers, Volume 7, Number 4, pp 291-303, April 2008.

[12] J. Gustafson "Reevaluating Amdahl's Law", Communications of the ACM, Volume 31, Number 5, pp 532-533, 1988.

[13] J. Liu, A. Vishnu, and D. K. Panda "Building Multirail InfiniBand Clusters: MPI-Level Design and Performance Evaluation", In Proceedings of the ACM/IEEE SC2004 Conference, pp 33 – 33, Nov. 2004.

[14] J. M. Squyres , "Processes, Processors, and MPI", Cluster World, MPI Mechanic Volume 1 Number 2, pp 8-11, January 2004.

[15] M. D. Hill and M. R. Marty, "Amdahl's Law in the Multicore Era", IEEE Computer Society, Volume 41, Number 7, pp 33-38, 2008.

[16] O. L. Atkin and D. J. Bernstein, "Prime sieves using binary quadratic forms", Mathematics of Computation Volume 73, pp 1023–1030, 2004.

[17] R. Brightwell, K. D. Underwood, "Evaluation of an Eager Protocol Optimization for MPI",10th European PVM/MPI Users' Group Meeting, Venice, Italy, pp 327-334, September 29 - October 2, 2003.

[18] R. M. Badia, J. Labarta, J. G., and F. Escal´e. DIMEMAS: Predicting MPI applications behavior in grid environments. In Workshop on Grid Applications and Programming Tools, 8th Global Grid Forum (GGF8), pages 50–60, Seattle, WA, June 2003.

[19] S. Gorlatch, "Send-Receive Considered Harmful: Myths and Realities of Message Passing", ACM Transactions on Programming Languages and Systems, Volume 26, Number 1, pp 47–56, January 2004.

[20] The MPI Forum. The MPI-2: Extensions to the Message Passing Interface, July 1997. Available at http://www.mpi-forum.org/docs/mpi-20-html/mpi2-report.html.

[21] V. S. Sunderam, "PVM: A framework for parallel distributed computing", Concurrency: Practice & Experience, Volume 2, Number 4, pp 315–339, Dec. 1990.

[22] W. Gropp, "MPICH2: A New Start for MPI Implementations", In Recent Advances in PVM and MPI: 9th European PVM/MPI Users' Group Meeting, Linz, Austria, Oct. 2002.

[23] Y. Aoyama J. Nakano "Practical MPI Programming", International Technical Support Organization, IBM Coorporation  SG24-5380-00, August 1999.

[24] Y. Yan, X. Zhang, and Q. Ma, "Software Support for Multiprocessor Latency Measurement and Evaluation", IEEE Transactions on Software Engineering, Volume 23, Number1, pp 4-16, January 1997.

## Author

**Alaa I. Elnashar** was born in Minia, Egypt, on November 5, 1967. He received his B.Sc. and M.Sc. from Faculty of Science, Department of Mathematics (Math. & Comp. Science), and Ph.D. from Faculty of Science, Department of Computer Science, Minia University, Egypt, in 1988, 1994 and 2005. He is a staff member in Faculty of Science, Computer Science Dept., Minia University, Egypt.

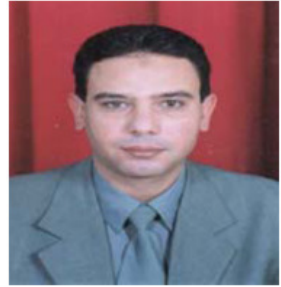

Dr. Elnashar was a postdoctoral fellow at Kanazawa University, Japan. His research interests are in the area of Software Engineering, Software Testing, Parallel programming and Genetic Algorithms.

Now, Dr Elnashar is an Assistant professor, Department of Computer Science, College of Computers and Information Technology, Taif University, Saudi Arabia